\documentclass[11pt]{article}
\usepackage{graphicx}
\usepackage{amsmath}
\usepackage{braket}
\usepackage{psfrag,epsfig}
\usepackage{subfigure}
\usepackage[percent]{overpic}
\usepackage{float}
\usepackage{color}

\usepackage{tikz}
\usetikzlibrary{arrows.meta,bending,positioning,intersections}
\usepackage{pgfplots}
\usepackage[all]{xy}
\pgfplotsset{compat=1.17}

\graphicspath{{figures/}}

\begin{document}

\title{Lubrication dynamics of a settling plate}
\author{Andrew Wilkinson,
Marc Pradas,
and Michael Wilkinson
\footnote{School of Mathematics and Statistics, The Open University, Walton Hall, Milton Keynes MK7 6AA, UK}
\footnote{email: m.wilkinson@open.ac.uk, marc.pradas@open.ac.uk}
}

\maketitle

\begin{abstract}
If a flat, horizontal, plate settles onto a flat surface, it is known 
that the gap $h$ decreases with time $t$ as a power-law: $h\sim t^{-1/2}$. 
We consider what happens if the plate is not initially horizontal, 
and/or the centre of mass is not symmetrically positioned: does 
one edge contact the surface in finite time, or does the plate 
approach the horizontal without making contact? The dynamics of this system is 
analysed and shown to be remarkably complex.
\end{abstract}

\section{Introduction}
\label{sec: 1}

We discuss a foundational problem in lubrication theory. 
We consider the motion of a body with a flat lower surface settling onto 
a flat horizontal surface, impeded by a thin layer of viscous fluid. For simplicity, 
we consider the two-dimensional case, where the gap $h$ at time $t$ depends 
only upon one Cartesian coordinate of the plane ($x$, say), and is independent 
of $y$. It is also assumed that the plate is sufficiently wide that fluid 
motion in the $y$-direction can be neglected. If the plate is initially horizontal,
and the weight acts through the centre, then it can be shown that the 
plate remains horizontal as it settles, with $h\sim t^{-1/2}$, in accordance with the 
classic lubrication result that smooth solid objects take an infinite amount 
of time to make contact~(\cite{Bre+61}).

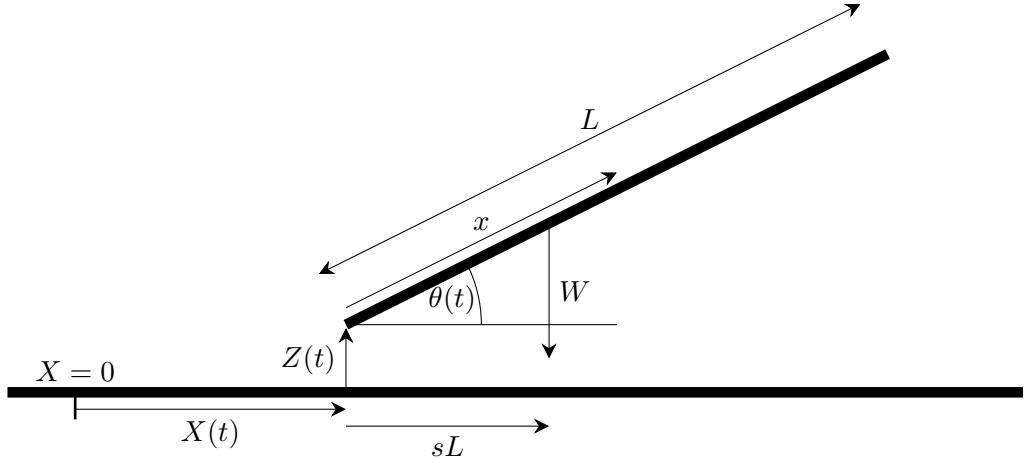
\begin{figure}
\centering
\begin{tikzpicture}[scale=0.9]
\path[draw,line width=4pt](2,0)--(17,0);
\path[draw,line width=4pt](7,1)--(15,5);
\draw[-{Stealth[length=2mm, width = 2mm]}] (7,0) -- (7,0.95) node[pos=.5,left] {$Z(t)$};
\draw[-] (7,1) -- (11,1) node[pos=.25,above] {$\ \ \ \ \ \ \ \ \theta(t)$};
\draw[-{Stealth[length=2mm, width = 2mm]}] (7,1.25) -- (11,3.25) node[pos=.5,above] {$x$};
\draw[-{Stealth[length=2mm, width = 2mm]}] (3,-0.25) -- (7,-0.25) node[pos=.5,below] {$X(t)$};
\draw[-{Stealth[length=2mm, width = 2mm]}] (10,2.5) -- (10,0.5) node[pos=.5,right] {$W$};
\draw[-{Stealth[length=2mm, width = 2mm]}] (7,-0.5) -- (10,-0.5) node[pos=.5,below] {$sL$};
\draw[{Stealth[length=2mm, width = 2mm]}-{Stealth[length=2mm, width = 2mm]}] (6.6,1.75) -- (14.6,5.74) node[pos=.5,above] {$L$};
\draw[line width=1pt] (3,-0.4)--(3,0) node [pos=1.0,above]{$X=0$};
\draw[] (9,1) arc (0:28:2);
\end{tikzpicture}
\caption{
The plate has width $L$ and its centre of mass is displaced from the left-hand edge by $sL$.
A coordinate $x$ measures distance from the left-hand edge, and all variables are assumed
to be independent of the other Cartesian coordinate in the plane, $y$.  
At a time $t$, the angle of the plate is $\theta(t)$, the gap at the left-hand edge is $Z(t)$ 
and the horizontal displacement of the left-hand edge is $X(t)$. 
}
\label{fig: 1}
\end{figure}

If the force is applied off-centre, and/or the plate is not initially horizontal, 
then the system is much harder to analyse. There are three coupled 
degrees of freedom: the gap $Z$ between the left-hand edge of the plate
and the surface, the angle of the plate $\theta$, and the horizontal 
displacement $X$ of the left hand edge (both $\theta$ and $Z$ must be assumed to
be small, in order for the lubrication theory approximations to be valid). 
The geometry of the system is illustrated in figure \ref{fig: 1}.

The equations of motion are the Reynolds equations of lubrication 
theory (\cite{Rey86}), reviewed in \cite{Mic50,Sze98}. 
The plate may, or may not, 
contact the surface in a finite time. We find that there are two different 
types of solution, those which the edge of the plate makes contact in finite time, 
and those where the gap decreases as a power-law. 

The motion of the plate is determined by three generalised forces: the vertical 
force $F_z$, the horizontal force $F_x$, and the torque (clockwise, about the left-hand edge), 
$G$. Section \ref{sec: 2} derives the equations of motion, by determining the resistance 
matrix which relates the force vector $(F_x,F_z,G)$ to the velocity vector, 
$(\dot X,\dot Z,\dot \theta)$. A similar expression for the resistance matrix
was obtained in \cite{Caw+10} for the problem of a falling wedge, but our equations do contain an 
additional term, which makes a significant difference to their solution. Section \ref{sec: 3} 
considers the reduction of the equations of motion to dimensionless form.

Sections \ref{sec: 4}, \ref{sec: 5} and \ref{sec: 6} discuss the solutions to these equations 
of motion. In section \ref{sec: 4} we show that the Reynolds lubrication equations
are exactly solvable, in terms of integrals of elementary functions, but that the 
general form of the solution is too complicated to be informative. This leads us to  
concentrate on understanding the qualitative form of the solutions, and their 
asymptotic approximations. We find that two qualitatively different types of solution exist. 
Section \ref{sec: 5} considers solutions for which $\theta/Z$ approaches a 
constant, using the dimensionless form of the equations of motion. 
Section \ref{sec: 6} discusses solutions where the plate contacts the 
surface on one edge in finite time. In this section we find it more convenient 
to use the original dimensional form of the equations. Section \ref{sec: 7} 
details some numerical experiments, comparing our solutions with numerical 
integration of the Reynolds equations. 
Section \ref{sec: 8} summarises our conclusions. 

\section{Equations of motion}
\label{sec: 2}

A flat plate of width $L$, immersed in a viscous fluid of viscosity $\mu$ and constant 
density $\rho$, settles onto a flat horizontal surface.  We consider the two-dimensional 
case, where the gap $h$ between both surfaces depends only upon one Cartesian 
coordinate of the plane ($x$, say), and is independent of $y$. It is also assumed that the plate is 
sufficiently wide that fluid  motion in the $y$-direction can be neglected.

The two surfaces move with velocities $v_1(t)$ (lower surface) and $v_2(t)$ (upper) and 
we assume that their  separation $h(x,t)$ is sufficiently small that lubrication 
theory applies (meaning that there is Poiseuille flow in the gap). The volume flux 
per unit depth is $J(x,t)$, the flux and pressure $p$ are related by
\begin{equation}
\label{eq: 2.1}
J=\left(\frac{v_1+v_2}{2}\right)h-\frac{h^3}{12\mu}\frac{\partial p}{\partial x}\, ,
\end{equation}
and the continuity equation  is 
\begin{equation}
\label{eq: 2.3}
\frac{\partial h}{\partial t}+\frac{\partial J}{\partial x}=0\, .
\end{equation}
Poiseuille flow also implies that the tangential stress on the upper surface is
\begin{equation}
\label{eq: 2.2}
\sigma=-\mu\left(\frac{v_2-v_1}{h}\right)-\frac{h}{2}\frac{\partial p}{\partial x}\, .
\end{equation}
We can use these equations in any convenient Cartesian frame. Rather than using the laboratory 
frame, let us use a frame which is attached to the plate. The displacement of the left-hand edge 
of the plate relative to some fixed position in the laboratory frame is $X(t)$. The velocity 
of the upper bounding surface in the plate frame is then $v_2=0$, and the velocity of the lower
surface is $v_1=-\dot X$.  The configuration of the 
plate is then described by specifying height at displacement $x$ from the left edge: 
\begin{equation}
\label{eq: 2.4}
h(x,t)=Z(t)+\theta(t)x,
\end{equation}
where $Z(t)$ is the gap between the left hand edge of the plate and the surface, and $\theta(t)$ 
is the angle between the plate and the horizontal, see Fig.~\ref{fig: 1}. 

We wish to determine the force components in the vertical and horizontal direction, $F_z$ and $F_x$, respectively, and torque $G$ on the plate. These are:
\begin{subequations}
\label{eq: Forces 1}
\begin{align}
F_z&=\int_0^L {\rm d}x\ p=p_0L-\int_0^L{\rm d}x\ x\frac{\partial p}{\partial x}\,, \\
F_x&=\int_0^L{\rm d}x\ \sigma -\theta F_z=-\mu\int_0^L{\rm d}x\ \frac{\dot X}{h}
-\int_0^L{\rm d}x\ \frac{h}{2}\frac{\partial p}{\partial x} -\theta F_z\,, \\
G&=\int_0^L{\rm d}x\ xp=-\int_0^L{\rm d}x\frac{x^2}{2}\frac{\partial p}{\partial x}
\,,
\end{align}
\end{subequations}
where $p_0 = p(0)$. The above expressions  depend on the pressure gradient $\partial p/\partial x$, which we can determine from Eq.~\eqref{eq: 2.1} as follows. Inserting Eq.~\eqref{eq: 2.4}  into the continuity equation \eqref{eq: 2.3} and integrating once we obtain
\begin{equation}
\label{eq: 2.10}
J=J_0-\dot Zx-\frac{\dot \theta}{2}x^2\,,
\end{equation}
where $J_0$ is a constant. It will be useful to define
\begin{equation}
\label{eq: 2.8}
I^n_m=\int_0^L{\rm d}x\ \frac{x^n}{h^m}
\,.
\end{equation}
We can now find the pressure gradient
\begin{equation}
\label{eq: 2.11}
\frac{\partial p}{\partial x}=12\mu\left[\dot Z\frac{x}{h^3}+
\frac{\dot \theta}{2}\frac{x^2}{h^3}-\frac{\dot X}{2}\frac{1}{h^2}-J_0\frac{1}{h^3}\right]
\,,
\end{equation}
where $J_0$ is found by integrating the above expression and imposing that $p(L)=p(0)$: 
\begin{equation}
\label{eq: 2.13}
J_0=\frac{1}{I^0_3}\left[\dot ZI^1_3+\frac{\dot \theta}{2}I^2_3-\frac{\dot X}{2}I^0_2\right]
\,.
\end{equation}
Provided the background pressure is high enough to prevent cavitation, its value 
must be irrelevant, so we set $p_0=0$. 
Therefore, the force components are:
\begin{subequations}
\label{eq: Forces 2}
\begin{align}
F_z=&\frac{12\mu}{I^0_3}\left[\dot X\frac{1}{2}\left(I^0_3I^1_2-I^0_2I^1_3\right)
+\dot Z\left(I^1_3I^1_3-I^2_3I^0_3\right)
+\dot \theta\frac{1}{2}\left(I^1_3I^2_3-I^3_3I^0_3\right)\right] \,, \\
F_x=&\frac{12\mu}{I^0_3}\left[\dot X\left(\frac{1}{6}I^0_1I^0_3-\frac{1}{4}I^0_2I^0_2\right)
+\dot Z\frac{1}{2}\left(I^1_3I^0_2-I^1_2I^0_3\right)
+\dot \theta\frac{1}{4}\left(I^0_2I^2_3-I^2_2I^0_3\right)
\right] 
\nonumber \\
&-\theta F_z\,, \\
G=&\frac{12\mu}{I^0_3}\left[\dot X\frac{1}{4}\left(I^2_2I^0_3-I^2_3I^0_2\right)
+\dot Z\frac{1}{2}\left(I^1_3I^2_3-I^3_3I^0_3\right)
+\dot\theta\frac{1}{4}\left(I^2_3I^2_3-I^4_3I^0_3\right)\right] 
\, .
\end{align}
\end{subequations}
We can write the relation between forces and velocities in terms of a matrix:
\begin{equation}
\label{eq: 2.17}
(F_x+\theta F_z,F_z,G)^{\rm T}=\frac{6\mu}{I^0_3}\, {\bf A}\, (\dot X,\dot Z, \dot \theta)^{\rm T}
\end{equation}
where the matrix ${\bf A}$ is
\begin{equation}
\label{eq: 2.18}
{\bf A}=
\left(\begin{array}{ccc}
\frac{1}{3}I^0_1I^0_3-\frac{1}{2}I^0_2I^0_2 
&
I^0_2I^1_3-I^1_2I^0_3
&
\frac{1}{2}\left(I^0_2I^2_3-I^2_2I^0_3\right)
\cr
I^1_2I^0_3-I^1_3I^0_2
&
2\left(I^1_3I^1_3-I^2_3I^0_3\right)
&
I^1_3I^2_3-I^3_3I^0_3
\cr
\frac{1}{2}\left(I^2_2I^0_3-I^0_2I^2_3\right)
&
I^1_3I^2_3-I^3_3I^0_3
&
\frac{1}{2}\left(I^2_3I^2_3-I^4_3I^0_3\right)
\end{array}
\right)
\ .
\end{equation}
Equations of motion in which a generalised velocity vector 
is obtained from a generalised force vector by matrix multiplication 
occur quite generally in treatments of viscosity dominated flow, and the 
matrix ${\bf A}$ is termed the \emph{resistance matrix} (\cite{Hap+83}).
The equations of motion (\ref{eq: 2.17}) and (\ref{eq: 2.18}) were obtained in a 
slightly different form in \cite{Caw+10}. In that work the contribution to the 
horizontal force which arises from tilting the plate was not included, and the 
first element of the force vector is $F_x$, rather than $F_x+\theta F_z$.

\section{Dimensionless equations}
\label {sec: 3}

It will be efficient to define non-dimensional dynamical variables:
\begin{equation}
\label{eq: 3.1}
\eta=\frac{L\theta}{Z}
\,,\ \ \ 
\zeta=\frac{X}{L}
\,,\ \ \ 
\xi=\frac{Z}{L} \,,
\end{equation}
and a non-dimensional version of the integrals $I^n_m$:
\begin{equation}
\label{eq: 3.2}
I^n_m=\frac{L^{n+1}}{Z^m}K^n_m(\eta)\,,
\end{equation}
with 
\begin{equation}
\label{eq: 3.3}
K^n_m(\eta)=\int_0^1 {\rm d}u\ \frac{u^n}{(1+\eta u)^m}
\,.
\end{equation}
With these definitions the generalised force components are expressed 
in terms of the dimensionless dynamical variables as follows:
\begin{subequations}
\label{eq: 3.4}
\begin{align}
F_z=&\frac{6\mu L}{K^0_3}\left[\frac{\dot \zeta}{\xi^2}(K^1_2K^0_3-K^1_3K^0_2)
+\frac{\dot \xi}{\xi^3}\left[(2K^1_3K^1_3-2K^0_3K^2_3)+\eta(K^1_3K^2_3-K^3_3K^0_3)\right]\right. \nonumber \\
&+\left.\frac{\dot \eta}{\xi^2}(K^1_3K^2_3-K^0_3K^3_3)
\right] \,, \\
F_x+\theta F_z=&\frac{6\mu L}{K^0_3}
\left[\frac{\dot \zeta}{\xi}(\frac{1}{3}K^0_1K^0_3-\frac{1}{2}K^0_2K^0_2)
+\frac{\dot \xi}{\xi^2}\left[(K^0_2K^1_3-K^1_2K^0_3)+\eta(\frac{1}{2}K^0_2K^2_3-\frac{1}{2}K^2_2K^0_3)
\right]\right. \nonumber \\
&+\left.\frac{\dot \eta}{\xi}(\frac{1}{2}K^0_2K^2_3-\frac{1}{2}K^2_2K^0_3)
\right]\,, \\
G=&\frac{6\mu L^2}{K^0_3}\left[
\frac{\dot \zeta}{\xi^2}(\frac{1}{2}K^2_2K^0_3-\frac{1}{2}K^2_3K^0_2)
+\frac{\dot \xi}{\xi^3}\left[(K^2_3K^1_3-K^3_3K^0_3)+\eta (\frac{1}{2}K^2_3K^2_3
-\frac{1}{2}K^4_3K^0_3)\right]\right. \nonumber \\
&+\left.\frac{\dot \eta}{\xi^2}(\frac{1}{2}K^2_3K^2_3-\frac{1}{2}K^4_3K^0_3)
\right]
\, .
\end{align}
\end{subequations}
It will also be helpful to introduce another dimensionless variable $\lambda$ 
which can be used in place of $\xi$, defined by:
\begin{equation}
\label{eq: 3.7}
\lambda=\ln \xi
\,.
\end{equation}
Then the force equations are
\begin{subequations}
\label{eq: 3.8}
\begin{align}
F_x&=\frac{\mu L}{\xi}\left[B_{11}(\eta)\dot \zeta+B_{12}(\eta)\dot \lambda+B_{13}(\eta)\dot \eta\right]\,, \\
F_z&=\frac{\mu L}{\xi^2}\left[B_{21}(\eta)\dot \zeta+B_{22}(\eta)\dot \lambda
+B_{23}(\eta)\dot \eta\right]\,,  \\
G&=\frac{\mu L^2}{\xi^2}\left[B_{31}(\eta)\dot \zeta+B_{32}(\eta)\dot \lambda
+B_{33}(\eta)\dot \eta\right]\,,
\end{align}
\end{subequations}
where the coefficients $B_{ij}(\eta)$ are obtained by comparison 
with (\ref{eq: 3.4}), e.g.
\begin{equation}
\label{eq: 3.9}
B_{11}(\eta)=\frac{2K^0_1K^0_3-3K^0_2K^0_2}{K^0_3}-6\eta \frac{K^1_2K^0_3-K^1_3K^0_2}{K^0_3}
\,.
\end{equation}
The equations of motion can then be expressed in the form
\begin{equation}
\label{eq: 3.10}
\frac{\xi^2}{\mu L}
\left(\begin{array}{c}
F_x/\xi \cr
F_z\cr
G/L
\end{array}\right)
={\bf B}
\left(\begin{array}{c}
\dot \zeta \cr
\dot \lambda \cr
\dot \eta
\end{array}\right)
\ .
\end{equation}
We are interested in settling of the body, due to a gravity force $W$ which acts through a point $x=sL$
from the left-hand edge, so the opposing forces on the body due to the fluid are 
$(F_x,F_z,G)=(0,W,sWL)$. Introduce a transformed time variable $\tau$, which satisfies
\begin{equation}
\label{eq: 3.11}
\frac{{\rm d}\tau}{{\rm d}t}=\frac{\xi^2W}{\mu L}
\end{equation}
so that, if the inverse of ${\bf B}(\eta)$ is ${\bf C}(\eta)$, then (\ref{eq: 3.10}) becomes
\begin{equation}
\label{eq: 3.12}
\left(\begin{array}{c}
\frac{{\rm d}\zeta}{{\rm d}\tau} \cr
\frac{{\rm d}\lambda }{{\rm d}\tau}\cr
\frac{{\rm d}\eta}{{\rm d}\tau}
\end{array}\right)
={\bf C}
\left(\begin{array}{c}
F_x /W \xi \cr
F_z/W \cr
G/WL
\end{array}\right)
=
{\bf C}\left(\begin{array}{c}
0 \cr
1 \cr
s
\end{array}\right)
\ .
\end{equation}
For our settling plate we then have the following equations of motion:
\begin{subequations}
\label{eq: 3.13}
\begin{align}
\frac{{\rm d}\zeta}{{\rm d}\tau}&=C_{12}(\eta)+sC_{13}(\eta)\equiv F_\zeta(\eta,s)\,, \label{eq: 3.13a}\\
\frac{{\rm d}\lambda}{{\rm d}\tau}&=C_{22}(\eta)+sC_{23}(\eta)\equiv F_\lambda(\eta,s)\,, \label{eq: 3.13b}\\
\frac{{\rm d}\eta}{{\rm d}\tau}&=C_{32}(\eta)+sC_{33}(\eta)\equiv F_\eta(\eta,s) 
\,. \label{eq: 3.13c}
\end{align}
\end{subequations}
The matrix elements of ${\bf B}(\eta)$  and ${\bf C}(\eta)$ can be determined analytically:
the expressions are given in Appendix A.

It is possible to find solutions in which $Z(t)$ is increasing, but the height of the centre 
of gravity must always decrease monotonically. It is not obvious that Equations 
(\ref{eq: 3.13a})-(\ref{eq: 3.13a}) have this property. Appendix B demonstrates that 
this property holds.

\section{Solution of dimensionless equations}
\label{sec: 4}

In subsection \ref{sec: 4.2} it will be shown that the equations of motion, 
(\ref{eq: 3.13}), can be 
solved, expressing $t$, $\xi$ and $\zeta$ as functions of $\eta$, defined as integrals 
of elementary functions. These exact solutions are, however, quite difficult to 
interpret. It is the qualitative behaviour of the solutions which is usually of more
interest than precise expressions. In particular, it is desirable to understand 
whether the edge of the plate makes contact in finite time, or whether the 
angle of the plate decreases so that it settles without ever making contact.
This is addressed in the next subsection \ref{sec: 4.1}, before we consider the exact solution.
 
\subsection{Fixed points}
\label{sec: 4.1}

The qualitative form of the solution  is addressed by looking at the 
dynamics of the dimensionless variable $\eta$. 
If the left-hand edge makes contact, this corresponds to $\eta\to \infty$ (and a 
contact of the right-hand edge is $\eta\to -1$).   

\begin{figure}
\centering
\includegraphics[width=0.99\textwidth]{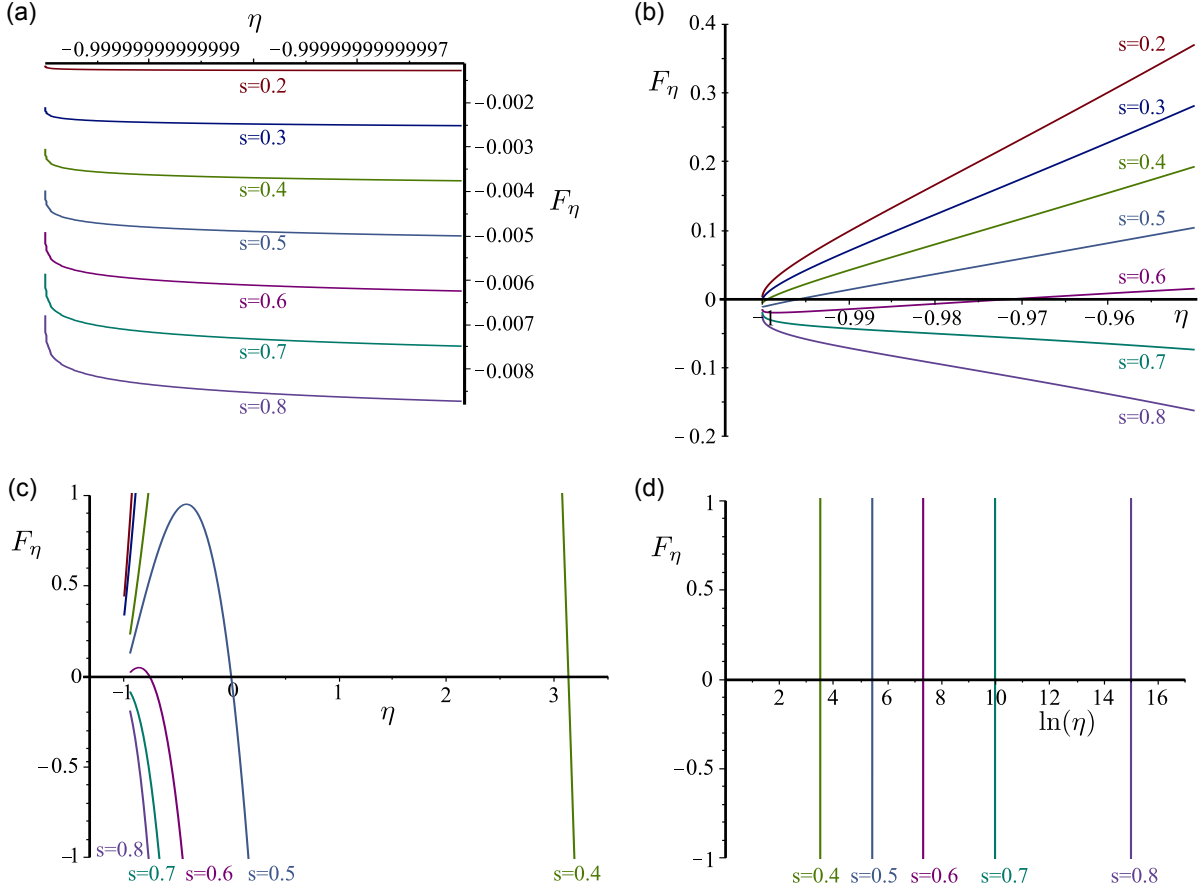}
\caption{
Plot of $F_\eta(\eta,s)$ as a function of $\eta$ for different choices of the centre of 
mass position, $s$, and for four different ranges of $\eta$. Attractive fixed points 
of $\eta$ occur where $F_\eta=0$ and $F'_\eta<0$. 
}
\label{fig: 2}
\end{figure}

Consider equation (\ref{eq: 3.13c}). This is a differential equation for $\eta(\tau)$ 
which is independent of the other variables. Any fixed points of the $\eta(\tau)$ dynamics 
are determined by the condition
\begin{equation}
\label{eq: 4.1}
F_\eta(\eta^\ast,s)=0
\, .
\end{equation}
This determines a fixed point $\eta^\ast(s)$ as a function of the position of the 
centre of gravity, $s$. The fixed point is stable if 
\begin{equation}
\label{eq: 4.2}
\kappa(s)\equiv -\frac{\partial F_\eta}{\partial \eta}(\eta^\ast(s),s)>0\,,
\end{equation}
(and, conversely, unstable if $\kappa <0$). 
The matrix elements quoted in 
equations (\ref{eq: A.2}) of Appendix \ref{appA} 
enable us to express $F_\eta(\eta,s)$ in terms of elementary functions: we find that 
the values of $s$ and $\eta$ at a fixed point are related by
\begin{equation}
\label{eq: 4.2a}
s=\frac {2\left(\ln \left( \eta+1 \right) \right)^2(\eta+1)+\ln \left( \eta+1\right)\eta(\eta+1)-3\eta^2}
 {\ln \left(\eta+1\right) \eta^2}
\ .
\end{equation}
Figure \ref{fig: 2} shows plots of $F_\eta(\eta,s)$ as a function of $\eta$ for different 
values of $s$. 
Because fixed points were found with  widely differing values of $\eta$, the plots 
of Fig.~\ref{fig: 2} correspond to four different ranges of $\eta$. Panel \ref{fig: 2}(a) illustrates the limit:
\begin{equation}
\label{eq: 4.2b}
\lim_{\eta\to -1}F_\eta(\eta,s)=0\,,
\end{equation}
i.e., $\eta=-1$ (right-hand edge approaches contact) is a fixed point for all $s$.
By symmetry, the limit $\eta\to \infty$ (i.e., left-hand edge approaches contact) 
can also be regarded as a fixed point. However, the function $F_\eta(\eta,s)$ 
has a singular form as $\eta\to -1$, and the dynamics in the vicinity of these 
edge-contact fixed points is highly unusual. In section \ref{sec: 6} we consider in 
some detail the case where the edge is close to contact.  There we argue that the 
plate makes contact in finite time.

Panels \ref{fig: 2}(b, c, d) show the behaviour of $F_\eta$ for three different 
ranges of $\eta$ as we increase its value above $\eta=-1$. We can see how fixed 
points are found at increasingly large values of $\eta$.
It would be instructive to plot a phase-diagram which shows the locations of the stable and 
unstable fixed points as lines in the $(\eta,s)$ plane, but the fact that $\eta$ may approach 
infinity is inconvenient. There is a left-right symmetry of the system, such that a fixed 
point at $(\eta,s)$ should be reflected as a fixed point with the same stability at 
$(-\eta/(1+\eta),1-s)$. If we were to use an alternative dimensionless parameter, 
\begin{equation}
\label{eq: 4.3}
\phi=\frac{\eta}{2+\eta}
\end{equation}
to describe the configuration of the plate, then the configuration space extends from 
$\phi=-1$ (right-hand edge in contact) to $\phi=+1$ (left-hand edge in contact), and 
a fixed-point at $(\phi,s)$ is mirrored by one at $(-\phi,1-s)$. In terms of the symmetrized 
configuration variable $\phi$, the equation for the fixed points is 
\begin{equation}
\label{eq: 4.3a}
s={\cal S}(\phi)\equiv 
\left(\frac{1-\phi^2}{2\phi^2}\right)\ln \left(\frac{1+\phi}{1-\phi}\right)
+\left(\frac{1+\phi}{2\phi}\right)
-\frac{3}{\ln \left(\frac{1+\phi}{1-\phi}\right)}
\,.
\end{equation}
The function ${\cal S}(\phi)$ satisfies the symmetry relation
\begin{equation}
\label{eq: 4.3b}
{\cal S}(\phi)+{\cal S}(-\phi)=1
\,.
\end{equation}

\begin{figure}
\centering
\includegraphics[width=0.49\textwidth]{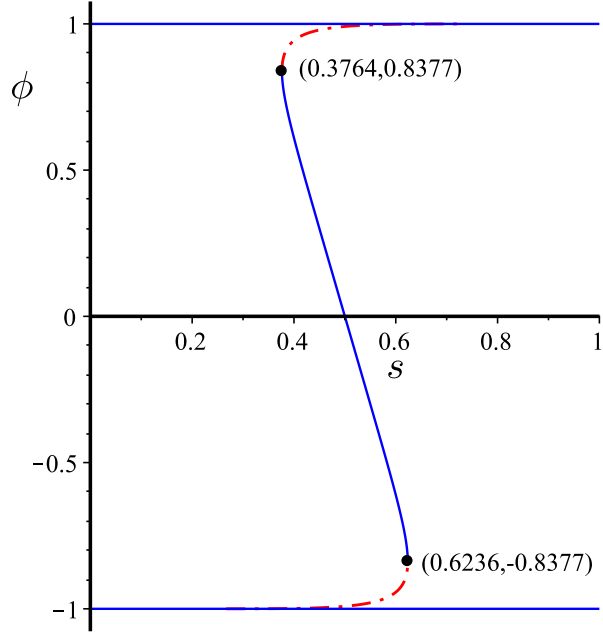}
\caption{
Plot showing fixed points in phase space of model, using $\phi$, defined by 
(\ref{eq: 4.3}) and (\ref{eq: 3.1}), to describe the configuration of the plate, 
and $s$ to parametrise the position of its centre of gravity. The blue line 
indicates stable fixed points, the red line unstable. 
}
\label{fig: 3}
\end{figure}

Figure \ref{fig: 3} shows the phase-diagram of the system using the $(\phi,s)$ 
coordinates. If we include the stable fixed points at $\phi=\pm 1$, 
there are either three or five fixed points, depending upon the value of $s$ (the center of mass position):

\begin{enumerate}

\item For $0<s<0.3764208755\ldots$ there is an unstable fixed point at a
value of $\phi^\ast$ which is very close to the point at which the right-hand side touches 
(note the dashed red line and solid horizontal blue lines are indistinguishable in Fig.~\ref{fig: 3}).
In this range of $s$, $\phi^\ast$ lies in $(-1,-0.999206\ldots]$. If the right-hand 
edge is exquisitely close to making contact, it does so. Otherwise, we expect 
that the left hand edge will make contact. In section \ref{sec: 6} we argue that, in both 
cases, contact happens in finite time.

\item For $1>s>0.623579125\ldots$, the situation is the mirror image of 
the case $0<s<0.3764208755\ldots$. There is an unstable fixed point 
at a value of $\phi^\ast$ which lies in the interval $[0.999206\ldots,1)$. If $\phi $ is initially 
greater than $\phi^\ast$, the left-hand edge contacts in finite time. If $\phi $ is 
less than $\phi^\ast$ (which is the only case which is of practical relevance), 
the right-hand edge contacts in finite time.

\item There is a region in the interval $0.3764208755\ldots\le s\le 0.6235791253\ldots$
which has two stable fixed points at $\phi=\pm 1$, two unstable fixed points, and 
one other stable fixed point between the unstable fixed points. 
There is a basin of attraction for which the value of $\phi$ 
approaches the non-trivial stable fixed point $\phi^\ast(s)$. This stable fixed point 
starts at $\phi^\ast(0.3764208755\ldots)=0.83774254\ldots$, passes
through $\phi^\ast(1/2)=0$, and disappears at 
$\phi^\ast(0.6235791253\ldots)=-0.83774254\ldots$. The basin of attraction 
in the $\phi$ variable is at its largest when $s=1/2$. At this point, 
the unstable fixed points are at $\phi=\pm 0.9956216$, and all initial conditions in 
between these values converge to the plate settling in the flat, $\phi=0$, configuration.
\end{enumerate}

In section \ref{sec: 5} we show that the approach to the fixed 
point has a power-law dependence upon time $t$.
  
\subsection{Exact solution}
\label{sec: 4.2}

We define a dimensionless time variable
\begin{equation}
\label{eq: 4.4}
\tilde t=\frac{Wt}{\mu L}\xi_0^2\,,
\end{equation}
where $\xi_0$ is the initial value of $Z/L$. 
Equations (\ref{eq: 3.11}) and (\ref{eq: 3.13}) 
then lead to the following differential equations for $\tilde t(\eta)$, $\lambda(\eta)$ and 
$\zeta(\eta)$:
\begin{subequations}
\begin{align}
\label{eq: 4.5}
\frac{{\rm d}\tilde t}{{\rm d}\eta}& =\xi_0^2\frac{\exp[-2\lambda(\eta,s)]}{F_\eta(\eta,s)}\,, \\
\label{eq: 4.6}
\frac{{\rm d}\lambda}{{\rm d}\eta}&=\frac{F_\lambda(\eta,s)}{F_\eta(\eta,s)}\,, \\
\label{eq: 4.7}
\frac{{\rm d}\zeta}{{\rm d}\eta}&=\frac{F_\zeta(\eta,s)}{F_\eta(\eta,s)}
\,.
\end{align}
\end{subequations}
These three differential equations can be integrated to express 
$t$, $\lambda=\ln \xi$ and $\zeta$ as functions of $\eta$, 
for example (\ref{eq: 4.5}) gives
\begin{equation}
\label{eq: 4.8}
t=\frac{\mu L}{W}\int_{\eta_0}^\eta {\rm d}\eta'\ 
\frac{\exp[-2\lambda(\eta')]}{C_{32}(\eta')+sC_{33}(\eta')}
\end{equation}
where $\lambda(\eta)$ is found by integrating \eqref{eq: 4.6}:
\begin{equation}
\label{eq: 4.9}
\lambda(\eta)-\lambda(\eta_0)=\int_{\eta_0}^\eta {\rm d}\eta'\ 
\frac{C_{22}(\eta')+sC_{23}(\eta')}{C_{32}(\eta')+sC_{33}(\eta')}
\ .
\end{equation}
These expressions give an exact expression for $t$ in terms of integrals 
of elementary functions of $\eta$, but they are too complicated to be 
instructive. Furthermore, in order to express $Z$, $\theta$ and $X$ as functions of $t$, 
it is necessary to invert the function $t(\eta)$ defined by (\ref{eq: 4.8}) and (\ref{eq: 4.9}).

\section{Asymptotic solution near stable fixed-point}
\label{sec: 5}

If the initial conditions are in the basin of attraction of a stable fixed point $\eta^\ast(s)$, 
then the long-time behaviour of the solution is determined by the solution in the vicinity
of this fixed point.

If $\eta$ is close to $\eta^\ast$ then 
\begin{equation}
\label{eq: 5.0}
\frac{d \eta}{d \tau} = F_\eta(\eta,s) \approx F_\eta(\eta^\ast,s) + 
(\eta - \eta^\ast) F'_\eta(\eta^\ast,s) = - (\eta - \eta^\ast) \kappa(s)
\,,
\end{equation}
so, with $\eta_0 = \eta(0)$, the solution of (\ref{eq: 3.13c}) is approximated by
\begin{equation}
\label{eq: 5.1}
\eta(\tau)\sim \eta^\ast+[\eta_0-\eta^\ast]\exp[-\kappa(s)\tau]
\,.
\end{equation}
where $\kappa(s)$ was defined in Equation \eqref{eq: 4.2}.
When $\eta(\tau)$ converges to this stable fixed point at $\eta^\ast$, the 
functions $F_\zeta$ and $F_\lambda$ in (\ref{eq: 3.13a}), (\ref{eq: 3.13b}) approach 
constant values. Equation (\ref{eq: 3.13b}) then implies that
\begin{equation}
\label{eq: 5.2}
\xi(\tau)\sim \xi_0 \exp[F_\lambda(\eta^\ast,s)\tau]\,,
\end{equation}
and hence equation (\ref{eq: 3.11}) can be integrated to give $t$ as a function of $\tau$ 
and $\xi_0$:
\begin{eqnarray}
\label{eq: 5.3}
t&= &\frac{\mu L}{W}\int_0^\tau {\rm d}\tau'\ \xi^{-2}(\tau')
\nonumber \\
&\sim&\frac{\mu L}{W\xi_0^2}\int_0^\tau {\rm d}\tau'\ \exp[-2F_\lambda(\eta^\ast,s)\tau']
\nonumber \\
&\sim&\frac{-\mu L}{2F_\lambda(\eta^\ast,s)W\xi_0^2}
\left(\exp[-2F_\lambda(\eta^\ast,s)\tau]-1\right)\,,
\end{eqnarray}
so that, (noting that (\ref{eq: 5.2}) implies that we must have $F_\lambda<0$), 
the leading term in the dependence of $\tau$ upon time is proportional to $\ln t$. 
Using (\ref{eq: 5.2}) in (\ref{eq: 5.3}) we obtain
\begin{equation}
\label{eq: 5.4} 
t\sim\frac{-\mu L}{2F_\lambda(\eta^\ast,s)W}
\left(\frac{1}{\xi^2(t)}-\frac{1}{\xi_0^2}\right)
\ .
\end{equation}
We can now determine dependences upon the true time, valid for $t\to \infty$. 
In the limit as $t\to \infty$, Equation \eqref{eq: 5.4} then implies
\begin{equation}
\label{eq: 5.5*}
\xi(t)\sim \sqrt{\frac{-\mu L}{2F_\lambda(\eta^\ast,s)W}}\ t^{-1/2}\,,
\end{equation}
and Equation \eqref{eq: 5.1} can be written as
\begin{equation}
\label{eq: 5.5}
\eta(t)\sim \eta^\ast+(\eta(0)-\eta^\ast)t^{-\gamma(s)}
\ ,\qquad
\gamma(s)=\frac{\kappa(s)}{2|F_\lambda(\eta^\ast,s)|}\,.
\end{equation}
The long-time evolution of $\zeta(t)$ is (assuming $X(0)=0$):
\begin{eqnarray}
\label{eq: 5.6}
\zeta(t)&\sim&F_\zeta(\eta^\ast,s) \tau
\nonumber \\
&\sim&\frac{F_\zeta(\eta^\ast,s)}{F_\lambda(\eta^\ast,s)}\ln\left(\frac{\xi(t)}{\xi_0}\right)
\nonumber \\
&\sim&K-\frac{F_\zeta(\eta^\ast,s)}{2F_\lambda(\eta^\ast,s)}\ln(t)
\nonumber \\
&\equiv&K-\tilde \zeta(s)\ln(t)\,,
\end{eqnarray}
(where $K$ and $\tilde \zeta$ are independent of $t$), so that the plate slides by an unbounded 
distance (unless $F_\zeta(\eta^\ast,s)=0$).

\begin{figure}
\centering
\includegraphics[width=0.95\textwidth]{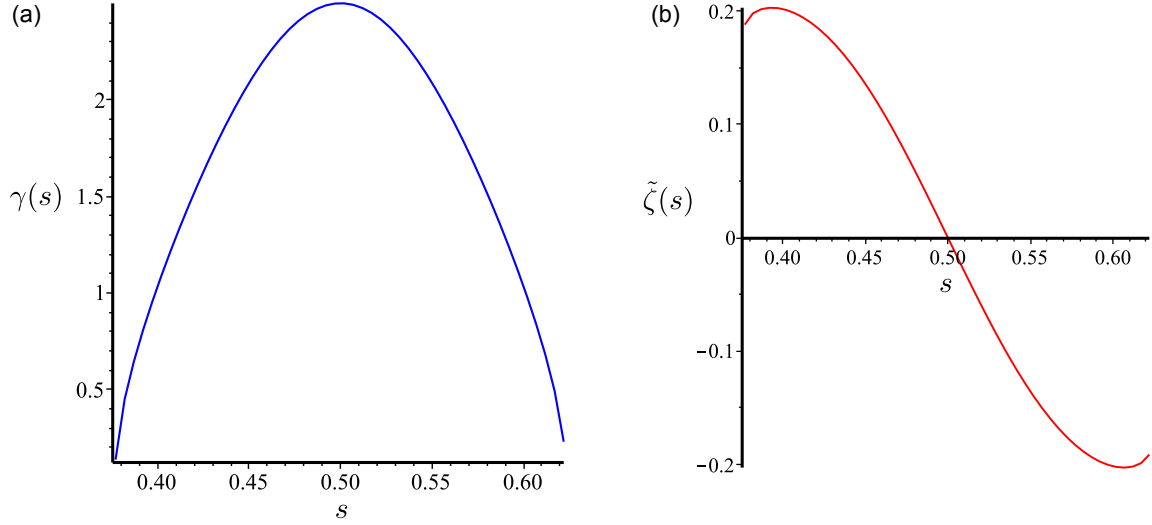}
\caption{
(a) Plot of the exponent $\gamma$, which characterises the relaxation of $\eta(t)$ towards its fixed 
point (equation (\ref{eq: 5.5})). (b) Coefficient of the logarithm, $\tilde \zeta$,
describing the slippage of the plate (equation (\ref{eq: 5.6})), as function of the 
centre of mass position parameter, $s$. 
}
\label{fig: 4}
\end{figure}

In the special case where the centre of mass is symmetrically placed, 
so that $s=0$, it is expected that the fixed point of (\ref{eq: 3.13c}) will be 
$\eta^\ast(1/2)=0$. So let us expand the matrix ${\bf C}$ about $\eta=0$. 
The leading order of $K^n_m(\eta)$ when $|\eta|\ll 1$ is:
\begin{equation}
\label{eq: 5.7}
K^n_m(\eta)=\int_{-1}^1{\rm d} x\ \frac{x^n}{(1+\eta x)^m}
\sim \int_{-1}^1{\rm d} x\ x^n(1-m\eta x)= \int_{-1}^1{\rm d} x\ x^n-m\eta \int_{-1}^1{\rm d} x\ x^{n+1}\,,
\end{equation}
so that 
\begin{equation}
\label{eq: 5.8}
K^n_m(\eta)\sim 
\left\{ \begin{array}{ll}
         \dfrac{2}{n+1}+\dfrac{2m(m+1)}{n+3}\eta^2 &\quad \mbox{for $n$ even}\,,\\
         & \\
        K^n_m(\eta)\sim \dfrac{-2m}{n+2}\eta &\quad  \mbox{for $n$ odd}.\end{array} \right. 
\end{equation}
Using the \emph{Maple} computer algebra system, we obtain up to second order:
\begin{equation}
\label{eq: 5.9}
{\bf C}=
\left(\begin{array}{ccc}
-1-\frac{1}{2}\eta+\frac{1}{12}\eta^2 
&
-\frac{1}{2}\eta-\frac{1}{4}\eta^2
&
0
\cr
\cr
-\frac{1}{2}\eta-\frac{1}{4}\eta^2
&
-16-18\eta-\frac{137}{35}\eta^2
&
30+39\eta+\frac{153}{14}\eta^2
\cr
\cr
\frac{1}{2}\eta^2
&
30+55\eta+\frac{405}{14}\eta^2
&
-60-120\eta-\frac{510}{7}\eta^2
\end{array}
\right)
\,,
\end{equation}
so that 
\begin{subequations}
\label{eq: 5.10}
\begin{align}
F_\zeta(\eta,1/2)&=-\frac{1}{2}\eta-\frac{1}{4}\eta^2+O(\eta^3)\,, \label{eq: 5.10a}\\
F_\lambda(\eta,1/2)&=-1+\frac{3}{2}\eta+\frac{31}{20}\eta^2+O(\eta^3)\,, \label{eq: 5.10b}\\
F_\eta(\eta,1/2)&=-5\eta-\frac{15}{2}\eta^2+O(\eta^3)
\,. \label{eq: 5.10c}
\end{align}
\end{subequations}
Hence $\kappa(1/2)=5$ and the exponent of $\eta$ for $s=1/2$ is
\begin{equation}
\label{eq: 5.11}
\gamma(1/2)=\frac{5}{2}\,.
\end{equation}
The motion of $X$ in the symmetric case is  different from Equation (\ref{eq: 5.6}) 
because $F_\zeta(0,1/2)=0$. In this case, using equation (\ref{eq: 5.1}) with 
$\tau \gg 1$, and $\eta_0 \ll 1$, equation (\ref{eq: 3.13a}) becomes
\begin{equation}
\label{eq: 5.12}
\frac{{\rm d}\zeta}{{\rm d}\tau}\sim-\frac{1}{2}\eta - 
\frac{1}{4}\eta^2 \sim \\-\frac{\eta_0}{2}\exp(-5\tau)
\, .
\end{equation}
Integrating this expression, we find that when $\eta\ll 1$ and $s=1/2$, 
$\zeta(\tau,\eta_0)\approx \frac{\eta_0}{10}[\exp(-5\tau)-1]$. 
In the limit as $\tau \to \infty$ and $\eta_0 \to 0$, the horizontal
displacement therefore remains finite:
\begin{equation}
\label{eq: 5.13}
\zeta\sim \zeta_0-\frac{\eta_0}{10}
\ .
\end{equation}

\section{Asymptotic solution near and following edge contact}
\label{sec: 6} 

In this section we first consider (section \ref{sec: 6.1}) the form of the solution of the 
lubrication equations when $\eta\gg 1$, so that the left-hand edge is very close to 
contact, showing that contact occurs after a finite time. 
We then consider (section \ref{sec: 6.2}) what happens after the left-hand edge has 
contacted the plate. Because this latter calculation has to start from first principles,
all of the calculations in this section will use the full, dimensional, equations of motion.
The results of section \ref{sec: 6.1} can be reproduced using the $\eta\to \infty$ 
limit of the coefficients in equations (\ref{eq: 3.13}). 

\subsection{Solution in the vicinity of left-edge contact}
\label{sec: 6.1}

The equation of motion for $\eta$ developed in section \ref{sec: 3} always has a 
stable fixed point at $\eta=-1$ (right-hand edge contacts), and by symmetry 
there must be a corresponding motion in which $\eta\to \infty$. Usually, a 
stable fixed point implies an exponential approach, but without the fixed point
being reached for any finite time. However, the form of the equations on the vicinity 
of $\eta=-1$ and $\eta\to\infty$ is so unusual that it is possible that 
$\eta\equiv L\theta/Z\to \infty$ in a finite time. In the following, we consider that possibility.

When $Z$ is very small, the integrals $I^n_m$ have simple asymptotic approximations:
\begin{equation}
\label{eq: 6.1}
I^n_m\sim \frac{1}{\theta^m}\int_{Z/\theta}^L{\rm d}v\ v^{n-m}\,,
\end{equation}
that is
\begin{equation}
\label{eq: 6.2}
I^n_m\sim \left\{
\begin{array}{cc}
\frac{1}{(k-1)Z^{k-1}\theta^{n+1}}&m=n+k\ ,\ \ k=2,3,\ldots \cr
&\cr
\frac{\ln(L\theta/Z)}{\theta^m}&m=n+1\cr
&\cr
\frac{L^{k+1}}{(k+1)\theta^m}&n=m+k\ ,\ \ k=0,1,\ldots
\end{array}
\right.
\ .
\end{equation}

Then, equations (\ref{eq:  3.4}) become
\begin{subequations}
%\label{eq: 6.3}
\begin{align*}
\frac{F_z}{24\mu\theta Z^2}&=
\frac{\dot X}{2}\left(\frac{\ln(L\theta/Z)}{2Z^2\theta^3}-\frac{1}{Z^2\theta^3}\right)
+\dot Z\left(\frac{1}{Z^2\theta^4}-\frac{\ln(L\theta/Z)}{2Z^2\theta^4}\right)
+\frac{\dot \theta}{2}\left(\frac{\ln(L\theta/Z)}{Z\theta^5}-\frac{L}{2Z^2\theta^4}\right) \,, \\
\frac{G}{24\mu\theta Z^2}&=
\frac{\dot X}{4}\left(\frac{L}{2Z^2\theta^3}-\frac{\ln(L\theta/Z)}{Z\theta^4}\right)
+\frac{\dot Z}{2}\left(\frac{\ln(L\theta/Z)}{Z\theta^5}-\frac{L}{2Z^2\theta^4}\right)
+\frac{\dot \theta}{4}\left(\frac{[\ln(L\theta/Z)]^2}{\theta^6}-\frac{L^2}{4Z^2\theta^4}\right)\,, \\
\frac{F_x+\theta F_z}{24\mu\theta Z^2}&=
\dot X\left(\frac{\ln(L\theta/Z)}{12Z^2\theta^2}-\frac{1}{4Z^2\theta^2}\right)
+\frac{\dot Z}{2}\left(\frac{1}{Z^2\theta^3}-\frac{\ln(L\theta/Z)}{2Z^2\theta^3}\right)
+\frac{\dot \theta}{4}\left(\frac{\ln(L\theta/Z)}{Z\theta^4}-\frac{L}{2Z^2\theta^3}\right)
\, . 
\end{align*}
\end{subequations}
We now set $F_x=0$, $F_z=W$, $G=WsL$, and retain leading order terms in 
the variable $L\theta/Z\equiv \eta$:
\begin{subequations}
\label{eq: 6.3}
\begin{align}
\frac{W}{6\mu}&=
\frac{\ln(L\theta/Z)}{\theta^2}\dot X
-\frac{2\ln(L\theta/Z)}{\theta^3}\dot Z
-\frac{L}{\theta^3}\dot \theta\,, \label{eq: 6.6}\\
\frac{WsL}{6\mu}&=
\frac{L}{2\theta^2}\dot X
-\frac{L}{\theta^3}\dot Z
-\frac{L^2}{4\theta^3}\dot \theta\,, \label{eq: 6.7}\\
\frac{W\theta}{6\mu}&=
\frac{\ln(L\theta/Z)}{3\theta}\dot X
-\frac{\ln(L\theta/Z)}{\theta^2}\dot Z
-\frac{L}{2\theta^2}\dot \theta \label{eq: 6.8}
\, . 
\end{align}
\end{subequations}
From Equations (\ref{eq: 6.6}) and (\ref{eq: 6.8}) we find
\begin{equation}
\label{eq: 6.10}
\dot X=-\frac{W\theta^2}{2\mu\ln(L\theta/Z)}
\ .
\end{equation}
Using the above equation with (\ref{eq: 6.7}) and (\ref{eq: 6.6}), and keeping only the 
leading order terms in $\eta=L\theta/Z$, we obtain
\begin{equation}
\label{eq: 6.11}
\dot Z=-\frac{1-s}{3}\frac{W}{\mu}\frac{\theta^3}{\ln(\eta)}
\, .
\end{equation}
Next substitute (\ref{eq: 6.10}) and (\ref{eq: 6.11}) into (\ref{eq: 6.7}) to obtain
\begin{equation}
\label{eq: 6.12}
\frac{\dot \theta}{\theta^3}=-\frac{2s}{3}\frac{W}{\mu L}
\,,
\end{equation}
which has solution
\begin{equation}
\label{eq: 6.13}
\theta(t)=\frac{\theta_0}{\sqrt{1+\frac{4s}{3}\theta_0^2Kt}}
\ ,\qquad
K=\frac{W}{\mu L}\,,
\end{equation}
where $\theta_0=\theta(0)$. Equations (\ref{eq: 6.10}), (\ref{eq: 6.11}) and (\ref{eq: 6.12}) are equations 
of motion for $X$, $Z$ and $\theta$, derived assuming that $\eta=L\theta/Z$ 
is large. We must consider whether (\ref{eq: 6.11}) and (\ref{eq: 6.12}) imply 
that $\eta\to \infty$ in finite time. Define a variable $K$ as in (\ref{eq: 6.13}), and a 
dimensionless pseudo-time $\tau$ (distinct from that defined by (\ref{eq: 3.11})) as follows:
\begin{equation}
\label{eq: 6.14}
\tau=Kt
\,.
\end{equation}
Noting that $\theta=\eta Z/L$, the equation of motion for $\theta$ is then
\begin{equation}
\label{eq: 6.15}
\frac{{\rm d}\theta}{{\rm d}\tau}=
\frac{1}{L}\left[Z\frac{{\rm d}\eta}{{\rm d}\tau}+\eta\frac{{\rm d}Z}{{\rm d}\tau}\right]
\,.
\end{equation}
Using (\ref{eq: 6.11}), and (\ref{eq: 6.13}), we obtain 
a differential equation for $\eta(\tau)$:
\begin{equation}
\label{eq: 6.16}
\frac{{\rm d}\eta}{{\rm d}\tau}=\frac{1-s}{3}\frac{\eta^2\theta^2}{\ln(\eta)}
-\frac{2s}{3}\eta \theta^2
\,.
\end{equation}
We now write $X=1+4s\theta_0^2 \tau/3$, and consider the limit as $\eta\to \infty$, 
where this equation of motion is approximated by 
\begin{equation}
\label{eq: 6.17}
X\frac{{\rm d}\eta}{{\rm d}X}=-\frac{4s}{1-s}\frac{\ln \eta}{\eta^2}\,,
\end{equation}
which has solution (with $\eta=\eta_0$ at $\tau=0$)
\begin{equation}
\label{eq: 6.18}
\ln\left[1+\frac{4}{3}s\theta_0^2\tau\right]=\frac{4s}{1-s}\
\left[\frac{\ln(\eta_0)+1}{\eta_0}-\frac{\ln(\eta)+1}{\eta}\right]
\,.
\end{equation}
When $\eta_0$ is very large, $\eta\to \infty$ in a finite time $\hat t$, which is approximated by
\begin{equation}
\label{eq: 6.19}
\hat t\sim \frac{3}{1-s}\frac{\ln (\eta_0)+1}{\eta_0\theta_0^2}\frac{\mu L}{W}
\,.
\end{equation}

\subsection{Motion after contact}
\label{sec: 6.2}

\begin{figure}
\centering
\begin{tikzpicture}[scale=0.9]
\path[draw,line width=1pt](2,0)--(10,0);
\path[draw,line width=1pt](8,3)--(16,3);
\path[draw,line width=1pt](2,0)--(8,3);
\path[draw,line width=1pt](10,0)--(16,3);
\draw[] (9,1) arc (0:180:0.1);
\draw[] (9-0.2,1) arc (0:-90:0.1);
\draw[] (9,1) arc (180:270:0.1);
\draw[] (12,2.6) arc (0:180:0.1);
\draw[] (12-0.2,2.6) arc (0:-90:0.1);
\draw[] (12,2.6) arc (180:270:0.1);
\draw[] (7.5,1.7) arc (0:180:0.1);
\draw[] (7.5-0.2,1.7) arc (0:-90:0.1);
\draw[] (7.5,1.7) arc (180:270:0.1);
\draw[] (4,0.3) arc (0:180:0.1);
\draw[] (4-0.2,0.3) arc (0:-90:0.1);
\draw[] (4,0.3) arc (180:270:0.1);
\draw[] (10,2.2) arc (0:180:0.1);
\draw[] (10-0.2,2.2) arc (0:-90:0.1);
\draw[] (10,2.2) arc (180:270:0.1);
\draw[] (10.5,0.7) arc (0:180:0.1);
\draw[] (10.5-0.2,0.7) arc (0:-90:0.1);
\draw[] (10.5,0.7) arc (180:270:0.1);
\draw[] (9,2.4) arc (0:180:0.1);
\draw[] (9-0.2,2.4) arc (0:-90:0.1);
\draw[] (9,2.4) arc (180:270:0.1);
\draw[] (7,0.4) arc (0:180:0.1);
\draw[] (7-0.2,0.4) arc (0:-90:0.1);
\draw[] (7,0.4) arc (180:270:0.1);
\end{tikzpicture}
\caption{
Schematic illustration of asperities on the surfaces. This simplified model 
assures that the pressure field is essentially unchanged across most of the 
surface, while the gap $h$ cannot be less than the prominence of the 
asperities, $\epsilon$.
}
\label{fig: 45}
\end{figure}
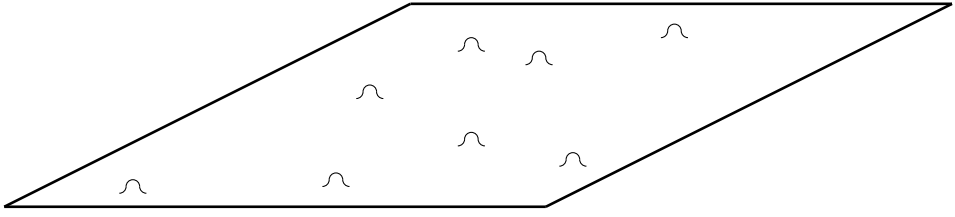

We have seen that, if the initial value of $\eta\equiv L\theta/Z$ is sufficiently large, 
the left hand edge contacts the surface in finite time. In order to complete our analysis 
of the problem, we need to consider what happens after contact is made. The motion 
after making contact depends upon the precise nature of the surfaces. 
The point of contact is able to tilt, and may also be able to slide. 
We assume that there are asperities on the contacting
surfaces, such that the gap $Z$ never falls below 
a microscopically small value, $\epsilon$. We assume that the typical separation 
between the asperities, $\Delta L$, satisfies $L\gg \Delta L\gg \epsilon$.
We also assume that $\theta$ satisfies $\theta \Delta L\ll \epsilon$, so that the size of the gap 
is nowhere reduced significantly due to tilt of the contacting surfaces.
If these asperities are widely-spaced, as illustrated in figure \ref{fig: 45},
the equations of motion for the generalised 
forces which are mediated by the pressure in the gap, namely $F_z$ and $G$, may 
be assumed to be obtained by a simple modification of equations (\ref{eq: 6.6}) 
and (\ref{eq: 6.7}): after contact occurs, $Z=\epsilon$, $\dot Z=0$, and there is an 
upward reaction force per unit depth $R$ at the left-hand edge. 

Because this additional force acts through the axis which 
defines the torque $G$, it makes no contribution to the torque equation. The modified 
forms of equations (\ref{eq: 6.6}) and (\ref{eq: 6.7}) are, therefore:
\begin{equation}
\label{eq: 6.20}
\frac{W-R}{6\mu}=
\frac{\ln(L\theta/\epsilon)}{\theta^2}\dot X
-\frac{L}{\theta^3}\dot \theta
\end{equation}
\begin{equation}
\label{eq: 6.21}
\frac{WsL}{6\mu}=
\frac{L}{2\theta^2}\dot X
-\frac{L^2}{4\theta^3}\dot \theta
\ .
\end{equation}
The equation for the horizontal force, $F_x$, will be changed more fundamentally because  
this component is sensitive to the precise form of the asperities. However, we shall 
not need the modified version of equation (\ref{eq: 6.8}).

Equations (\ref{eq: 6.20}) and (\ref{eq: 6.21}) imply that 
\begin{equation}
\label{eq: 6.22}
\dot X=\frac{L\theta^2}{\ln\left(\frac{L\theta}{\epsilon}\right)}\left[\frac{W-R}{6\mu L}
+\frac{\dot \theta}{\theta^3}\right]\,,
\end{equation}
so that the term proportional to $\dot X$ in (\ref{eq: 6.21}) is negligible as 
$\epsilon \to 0$. After contact, the equation of motion for $\theta$ is
very well approximated by
\begin{equation}
\label{eq: 6.23}
\frac{\dot \theta}{\theta^3}=-4s\frac{W}{6\mu L}\,,
\end{equation}
and (\ref{eq: 6.22}) implies that the sliding velocity, $\dot X$, while being indeterminate 
because we have not determined the reaction force $R$, is very small. Combining 
(\ref{eq: 6.22}) and (\ref{eq: 6.23}), we find a relation between $\dot X$ and $R$:
\begin{equation}
\label{eq: 6.24}
\dot X=\frac{\theta^2}{6\mu \ln\left(\frac{L\theta}{\epsilon}\right)}\left[W(1-4s)-R\right]
\,.
\end{equation}
Note that the equation of motion for $\theta$, (\ref{eq: 6.23}) is the same as 
that which is applicable before contact (equation (\ref{eq: 6.12})), whereas 
(\ref{eq: 6.22}) implies that the value of $\dot X$ changes upon contact. 
The values of $\dot X$ and $R$ remain ambiguous because we have 
not specified how the equation for the horizontal force is modified when 
surfaces are in contact. We now assume 
that, whenever there is a non-zero reaction force between the plate and the surface, the 
plate is prevented from moving horizontally, so that $\dot X=0$ when $R>0$. 
Then equation (\ref{eq: 6.24}) implies that the reaction force is 
\begin{equation}
\label{eq: 6.24}
R=(1-4s)W
\,.
\end{equation}
The condition that the reaction force $R$ must be non-negative cannot, therefore, 
be satisfied, if $s>1/4$ and $\dot X=0$, because the torque exerted by the downward force
causes the left-hand edge of the plate to lift. If $s>1/4$, the reaction force $R$ is reduced to 
zero, and the plate is then able to slide. The sliding motion produces additional contributions
to the force $F_z$ and the torque $G$, such that the left-hand edge remains in contact 
as it slides with reaction force $R=0$.
By symmetry, there must be sliding motion if $1/4<s<3/4$. 
In this range of $s$, the slow sliding motion creates a downward force 
on the plate which counters the effect of the torque. 
After contact the plate can slide over the surface so that the reaction force is $R=0$.
The slip velocity when the left-hand edge is in contact is
\begin{equation}
\label{eq: 6.26}
\dot X=(1-4s)\frac{W\theta^2}{6\mu \ln\left(\frac{L\theta}{\epsilon}\right)}
\,.
\end{equation} 

\section{Numerical studies}
\label{sec: 7}

We have compared our analytical predictions with numerical solutions of the equations
of motion. In each case we compared the numerical solution with the exact solution, and with 
the appropriate asymptotic approximation. We used dimensionless variables 
in all of the simulations, equivalent to setting $L=\mu=\rho=W=1$.

The exact solution given in subsection \ref{sec: 4.2} was evaluated as follows.
First we convert the initial values of the coordinates $(X,Z,\theta)$ to the dimensionless
coordinates $(\zeta,\lambda,\eta)$. We then integrated equations (\ref{eq: 4.5}), (\ref{eq: 4.6})
and (\ref{eq: 4.7}) to obtain $\tilde t$, $\lambda $ and $\zeta$ as functions of $\eta$. 
In the case where we expect convergence to a stable fixed point of $\eta$, these integrations 
can only be continued to a point close to the fixed-point value. 
The values of $\tilde t$ were stored in a table along with the values 
of $\zeta$, $\lambda$ and $\eta$. The plots of the exact solution are generated by 
interpolating this table to plot $\eta$, $\zeta$ or $\xi=\exp(\lambda)$ as a function of $\tilde t$.

\begin{figure}
\centering
\includegraphics[width=0.92\textwidth]{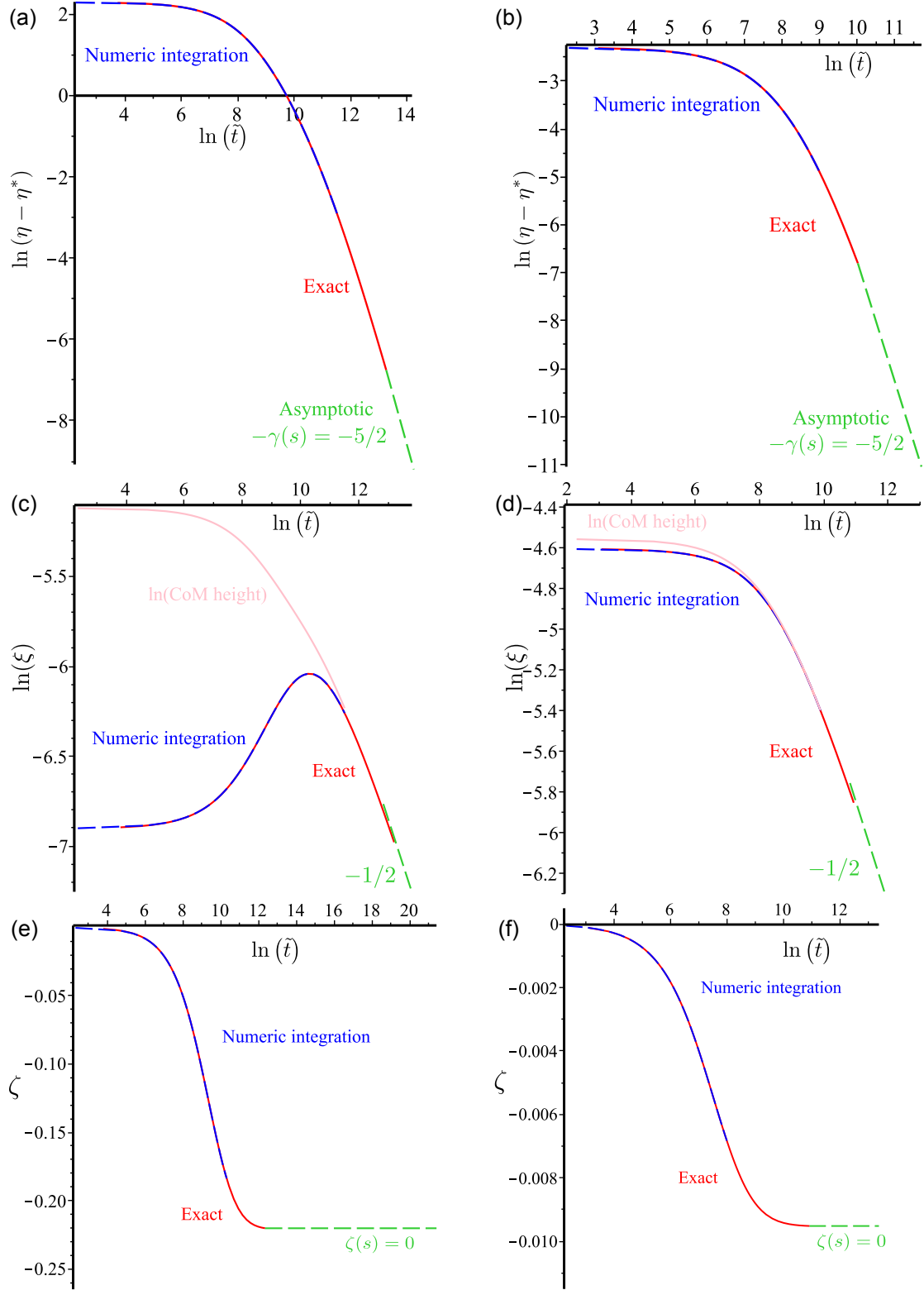}
\caption{
\label{fig: 6}
Comparison of numerical solutions of the equations of motion with analytical 
predictions for the case where the centre of gravity is symmetrically placed, $s=1/2$, 
and when the parameter $\eta(t)=L\theta(t)/Z(t)$ approaches a stable
fixed point.  Panels (a,c,e) and (b,d,f) correspond to two different sets of initial 
conditions. Both $\eta(\tilde t)$ (a,b) and the dimensionless plate height 
$\xi(\tilde t)=Z(\tilde t)/L$ (c,d)  show a power-law relaxation at large $t$, 
and the horizontal displacement $\zeta(\tilde t)=X(\tilde t)/L$ (e,f) approaches  a constant.}
\end{figure}

\begin{figure}
\centering
\includegraphics[width=0.92\textwidth]{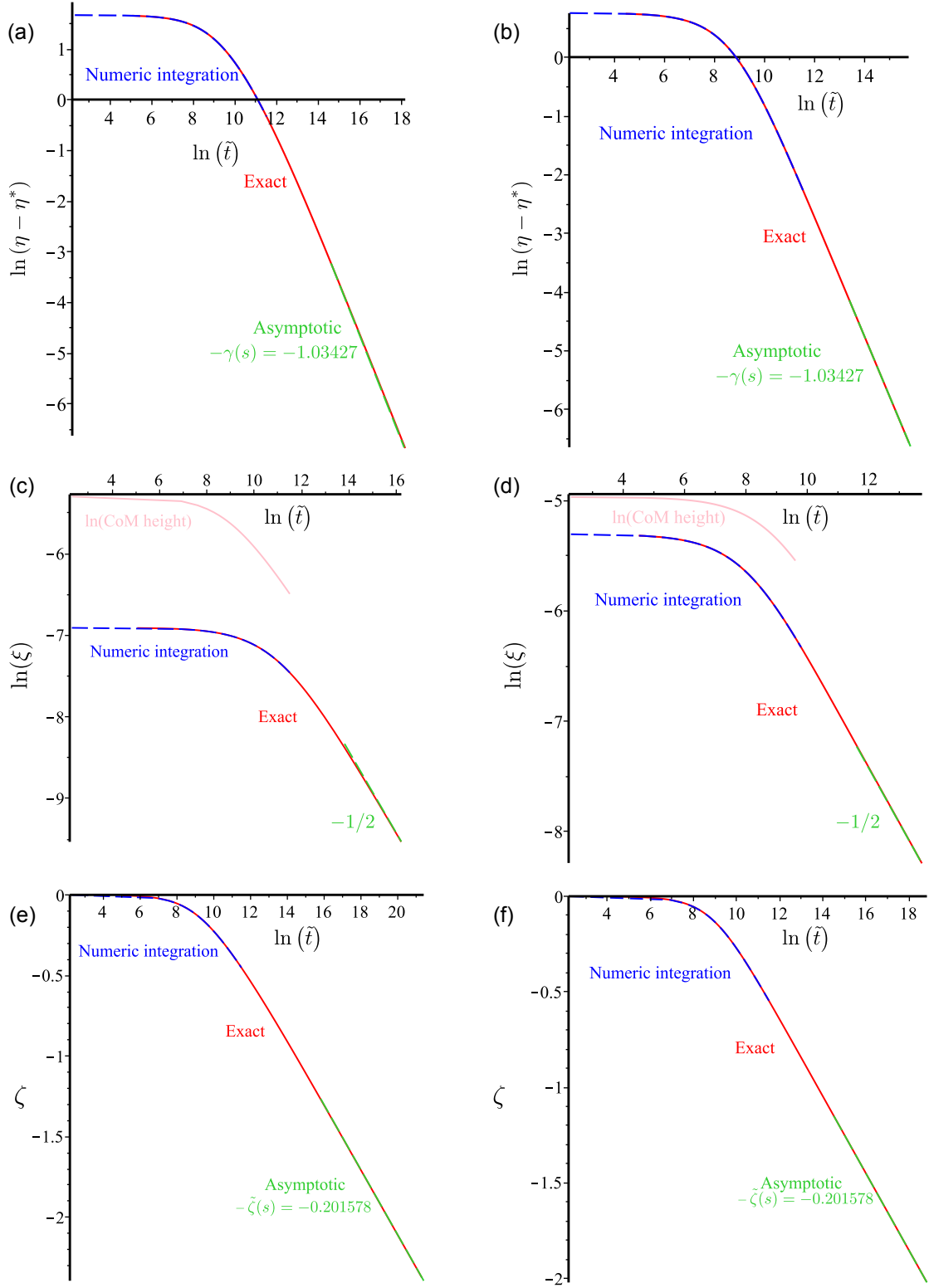}
\caption{
\label{fig: 7}
Comparison of numerical solutions of the equations of motion with analytical 
predictions for the case where the centre of gravity is displaced from the center, 
$s=0.4$, and when the parameter $\eta(t)=L\theta(t)/Z(t)$ approaches a stable
fixed point.  Panels (a,c,e) and (b,d,f) correspond to two different sets of initial 
conditions. The exponent for the relaxation of $\eta $, given by Equation (\ref{eq: 5.5}), 
is different than in the symmetric case [compare panels (a,b) with Figure \ref{fig: 6}(a,b)], 
but $\xi(\tilde t)=Z(\tilde t)/L\sim \tilde t^{-1/2}$  continues to hold in this asymmetric 
case (c,d). The horizontal  displacement $\zeta(\tilde t)=X(\tilde t)/L$ (e,f) grows 
logarithmically, as given by Equation (\ref{eq: 5.6}).}
\end{figure}

\begin{figure}
\centering
\includegraphics[width=0.94\textwidth]{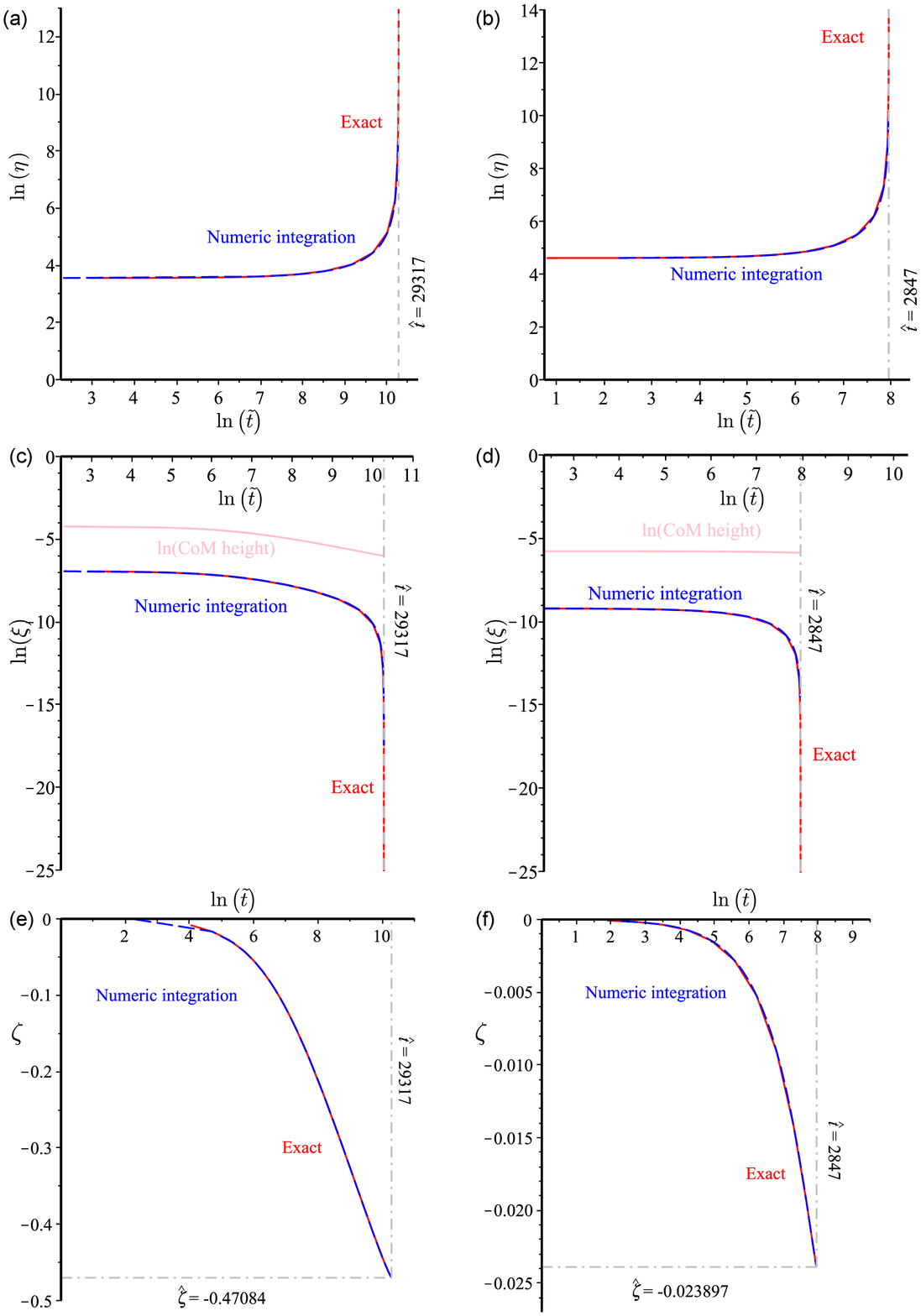}
\caption{
\label{fig: 8}
Plot of trajectory which makes contact in finite time, showing 
$\eta(\tilde t)$, $\xi(\tilde t)$ and $\zeta(\tilde t)$ up to the point of contact.
({\bf a,c,e}) $s=0.4$, $(X,Z,\theta)=(0,0.001,0.035)$, for which $\eta=35$ 
(which is beyond the unstable fixed point at $\eta=33.20746$, so outside 
the basin of attraction of the non-trivial stable fixed point at $\eta=3.13266$).
({\bf b,d,f}) $s=0.3$, $(X,Z,\theta)=(0,0.0001,0.01)$.}
\end{figure}

Figures \ref{fig: 6} and \ref{fig: 7} illustrate motion in which $\eta$ converges to a 
stable fixed point. 

Figure \ref{fig: 6}, shows the case where the centre of gravity is symmetrically placed, 
$s=1/2$, so that the stable fixed point is at $\eta^\ast=0$ 
(equivalently $\phi^\ast=0$), and there are unstable fixed points at $\eta=-0.99562$ 
and $\eta=227.39933$. We consider two different sets 
of initial conditions which are: $(X,Z,\theta)=(0,0.001,0.01)$
and $(X,Z,\theta)=(0,0.01,0.001)$, corresponding to initial values 
$\eta=10$, ($\phi=0.83333$) and $\eta=0.1$, ($\phi=0.047619$) respectively. 
We plot $\eta(\tilde t)$ and $\xi(\tilde t)$ on double-logarithmic scales, showing the 
asymptotic power-law behaviour of these functions, with exponents 
$-\gamma(1/2)=-5/2$ and $-1/2$ respectively, in agreement with equations (\ref{eq: 5.11}) 
and (\ref{eq: 5.5*}). The second set of initial conditions also show the finite 
horizontal plate movement, close to $-\eta_0/10$ ($=-0.01$ here), as expected 
from equation (\ref{eq: 5.13}). For the first set of initial conditions, $Z(t)$ initially 
increases, but in both cases the height of the centre of mass always decreases 
monotonically.

Figure \ref{fig: 7} considers an asymmetric case, where $s=0.4$, which has a stable 
fixed point at $\eta^\ast=3.13266$ (or $\phi^\ast=0.61033$), and there are unstable 
fixed points at $\eta=-0.99933$ and $\eta=33.20746$ . We consider two different sets 
of initial conditions which are: $(X,Z,\theta)=(0,0.001,0.01)$
and $(X,Z,\theta)=(0,0.005,0.005)$, corresponding to initial values $\eta=10$, ($\phi=0.83333$)  
and $\eta=1$, ($\phi=0.33333$) respectively. Both initial conditions are within the 
basin of attraction of the stable fixed point. 
The double-logarithmic scale plot of $\xi(\tilde t)$ shows the asymptotic 
power-law behaviour with exponent $-1/2$ also holds in the asymmetric case, but
the exponent $\gamma(s)$ for the decrease of $\eta(t)$ takes a different value, plotted 
in figure \ref{fig: 4}(a). In figure \ref{fig: 7}, the horizontal displacement is plotted as a
function of $\ln \tilde t$, in accord with the prediction that $X(t)$ increases logarithmically, 
with coefficient plotted in figure \ref{fig: 4}(b).

Figure \ref{fig: 8} illustrates cases in which the left-hand edge 
of the plate makes contact with the surface, corresponding to $\eta\to \infty$.
We show trajectories for the following cases ($L=1$ throughout): 
\begin{enumerate}
\item[(i)] (Panels a, c, e) $s=0.4$, $(X,Z,\theta)=(0,0.001,0.035)$, for which 
$\eta=35$ ($\phi=0.94595$), beyond the unstable fixed point at $\eta=33.20746$.
\item[(ii)] (Panels  b, d, f) $s=0.3$, 
$(X,Z,\theta)=(0,0.0001,0.01)$, for which $\eta=100$ ($\phi=0.98039$) 
and there is an unstable fixed point at $\eta=-0.99995$ and the only stable 
fixed points are at $\eta=-1$ and $\eta \to \infty$. 
\end{enumerate}
In both cases we plot $\eta(\tilde t)$, $\xi(\tilde t)$ and $\zeta(\tilde t)$ 
up to the point of contact. The contact time, $\hat t$, is shown in each
case.  For the first set of initial conditions, $\eta_0$ is close to 
a fixed point, so that $\hat t$ is much larger than the prediction of 
Equation (\ref{eq: 6.19}). Our second case has a relatively large initial $\eta$ value of $100$ and 
here the contact time is $\hat t=2847$, approaching the theoretical contact time of 
$2402$  predicted by equation (\ref{eq: 6.19}) for very large initial $\eta$ values.

\section{Concluding remarks}
\label{sec: 8}

Our motivation for examining the lubrication theory for a 
settling plate came from the observation that there 
are two plausible types of solution: the plate might \lq side-slip' 
onto the surface, making contact in finite time, or the motion 
might \lq round out', so that the plate settles in a progressively 
flatter attitude, without ever making contact.

We find that both of these types of solution may be realised, depending upon the 
initial position of the plate and the position of the centre of force.
We find that there is only one dimensionless coordinate which controls 
the nature of the trajectory: throughout most of the calculations we used 
$\eta=L\theta/Z$, but if we wish to represent the phase diagram of the system 
(Figure \ref{fig: 3}) in a manner which reflects its symmetry, the alternative choice
$\phi=\eta/(2+\eta)$ can be used. 

The elements of both the resistance matrix (detailed in the Appendix) 
and its inverse can be expressed 
exactly as elementary functions of $\eta$. This observation was used to write an 
exact solution of the equations of motion, in parametric form, in terms of integrals 
of elementary functions. The exact solution 
(presented in section \ref{sec: 4.2}) is, however, difficult to interpret. 
This led us to investigating asymptotic approximations to the solution.
We investigated two types of asymptotic solution:

\begin{enumerate}

\item There are solutions where $\eta$ approaches a constant $\eta^\ast$, with both 
$Z$ and $\theta$ decreasing as $t^{-1/2}$. Numerical experiments show that 
these solutions only exist for $s\in [s^\ast,1-s^\ast]$, with $s^\ast\approx 0.376\ldots $.
The value of $X(t)$ shows a logarithmic evolution, except for the case $s=1/2$, and 
the relaxation of $\eta(t)$ towards $\eta^\ast$ is a power-law, $\eta-\eta^\ast\sim t^{-\gamma(s)}$.
The exponent $\gamma(s)$ has a maximum $\gamma(1/2)=5/2$, and approaches zero as 
$s\to s^\ast$ or $s\to 1-s^\ast$.

\item For all values of $s$ there are also solutions where either of the edges will contact 
the baseplate in finite time, although the basin of attraction of one or both of these solutions
may be extremely small. For $s\in[0,1/4]$ or $s\in[3/4,1]$, there is a solution where the plate 
contacts without sliding ($Z=0$, $\dot X=0$). For $s\in[1/4,3/4]$, however, 
the plate must slip slowly in order to maintain contact. In both cases (whether or not the point 
of contact slips), the angle decreases as $\theta\sim t^{-1/2}$.

\end{enumerate}

Declaration of interests. {The authors report no conflict of interest.

\section*{Appendix A: Matrix coefficients}

We list the matrix elements of ${\bf B}$, defined by equation (\ref{eq: 3.8}).
In the following expressions, $\psi =\ln(\eta +1)$. The elements are:
\begin{subequations}
\label{eq: A.1}
\begin{align}
B_{11}&= {\frac { \left( -4\eta-8 \right) \psi +6\eta}{ \left( \eta+2 \right) \eta}}\\
B_{12}&={\frac { \left( -6\eta-6 \right) \psi +3\eta^2+6\eta}{ \left( \eta+2 \right) {\eta}^{2}}}\\
B_{13}&={\frac { \left( -12\,\eta-18 \right) \psi +3
\,{\eta}^{2}+18\,\eta}{ \left( \eta+2 \right) {\eta}^{3}}}\\
B_{21}&={\frac { \left( 6\,\eta+12 \right) \psi -12
\,\eta}{ \left( \eta+2 \right) {\eta}^{2}}} \\
B_{22}&={\frac { \left( 12\,\eta+12 \right) \psi -6
\,{\eta}^{2}-12\,\eta}{ \left( \eta+2 \right) {\eta}^{3}}} \\
B_{23}&= {\frac { \left( 24\,\eta+36 \right) \ln  \left( \eta+1 \right) -6
\,{\eta}^{2}-36\,\eta}{ \left( \eta+2 \right) {\eta}^{4}}} \\
B_{31}&= {\frac { \left( -12\,\eta-18 \right) \psi+3
\,{\eta}^{2}+18\eta}{ \left( \eta+2 \right) {\eta}^{3}}}\\
B_{32}&= -\frac32\,{\frac { \left(  \left( 2\eta+2 \right) \psi +{\eta}^{2}-2\,\eta \right)  \left(  \left( -2\eta-2
 \right) \psi+{\eta}^{2}+2\eta \right) 
}{ \left( \eta+2 \right) {\eta}^{5}}}\\
B_{33}&= {\frac {12\, \left( \eta+1 \right) ^{2} \psi^{2}+ \left( -72{\eta}^{2}-96\,\eta
 \right) \psi-3\,{\eta}^{2} \left( {\eta}
^{2}-4\,\eta-28 \right) }{ 2\left( \eta+2 \right) {\eta}^{6}}}
\end{align}
\end{subequations}
The matrix elements of ${\bf C}={\bf B}^{-1}$ are:
\begin{subequations}
\label{eq: A.2}
\begin{align}
C_{11}&=-\frac {\eta}{\psi}\\
C_{12}&={\frac {-{\eta}^{2}}{2\psi}} \\
C_{13}&=0 \\
C_{21}&= {\frac {-{\eta}^{2}}{2\psi}}\\
C_{22}&=\frac{\left(4\, \left( \eta+1 \right)^{2} \psi^{3}-\left(6\eta^{2}+8\eta \right) \psi^{2}-\eta^{2} \left(\eta^2+8\eta+2 \right) \psi +3\eta^{4}+6\eta^{3} \right) {\eta}^{3}}
{3 \left(  \left( -2\eta-2 \right) \psi +{\eta}^{2}+2\eta \right) \psi  \left( 
 \left( 2\eta+2 \right) \psi^{2}+ \left( {\eta}^{2}+2\eta \right) \psi -4\eta^{2} \right) }\\
C_{23}&=\frac {\left(\left( -4\eta-6 \right) \psi+\eta^2+6\eta \right) \eta^5}
 {3((-2\eta-2)\psi+\eta^2+2\eta)  (( 2\eta+2) \psi^2+ (\eta^2+2\eta) \psi-4\eta^2)}\\
C_{31}&={\frac {{\eta}^{3}}{2\psi}} \\
C_{32}&={\frac { \left(  \left( 2\eta+2 \right)  \psi^{2}+ \left( {\eta}^{2}+\eta \right) 
\psi-3\eta^{2} \right) {\eta}^{4}}
{3\psi \left(  \left( 2\eta+2 \right) \psi ^{2}+ \left( {\eta}^{2
}+2\,\eta \right) \psi-4\eta^{2}
 \right) }}\\
C_{33}&=
-{\frac {{\eta}^{6}}{ \left( 6\,\eta+6 \right)  \psi^{2}+ \left( 3\eta^{2}+6\eta \right) \psi-12\eta^{2}}}
\end{align}
\end{subequations}

\section*{Appendix B: Demonstration that system is always dissipative}

The height of the centre of gravity is 
\begin{equation}
\label{eq: B.1}
Z_{\rm c}=Z+sL\theta\equiv L\xi_{\rm c}
\end{equation}
where $\xi_{\rm c}=\xi(1+s\eta)$. We can also define $\lambda_{\rm c}\equiv\ln \xi_{\rm c}$, 
and $Z_{\rm c}$ is always decreasing if $\lambda_{\rm c}$ is decreasing as a function of time
$t$. Noting that the pseudo-time $\tau$ (defined by (\ref{eq: 3.11})) is a 
monotonically increasing function of time $t$,
it suffices to show that
\begin{equation}
\label{eq: B.2}
\frac{{\rm d}\lambda_{\rm c}}{{\rm d}\tau}=\frac{{\rm d}\lambda}{{\rm d}\tau}
+\frac{s}{1+s\eta}\frac{{\rm d}\eta}{{\rm d}\tau}<0
\,.
\end{equation}
Using the equations of motion (\ref{eq: 3.13b}), (\ref{eq: 3.13c}), this can be 
expressed as an inequality involving the matrix elements of ${\bf C}$:
\begin{equation}
\label{eq: B.3}
K(\eta,s)\equiv (1+s\eta)\left[C_{22}(\eta)+sC_{23}(\eta)\right]
+s\left[C_{32}(\eta)+sC_{33}(\eta)\right]<0
\end{equation}
This argument shows that, for any $s$, the inequality ensuring that the centre of 
mass falls is expressed in terms 
of just one variable, $\eta$. 

The definitions of the dimensionless variables $\xi$ and $\eta$ are based 
upon the separation $Z$ at the left-hand edge of the plate. It is sufficient to 
consider only the case where gap at the left-hand edge is smaller than or 
equal to that at the right-hand edge, that is, $\eta\ge 0$. If we are interested 
in the case where the right-hand gap is smaller, we can apply a reflection, and change
the centre of gravity parameter from $s$ to $1-s$. So it is sufficient to show that 
the maximal values of $K(\eta,s)$ and $K(\eta,1-s)$ for $\eta\ge 0$ are both negative.

Rather than using analysis to prove that this inequality holds, 
we provide a numerical demonstration. Figure \ref{fig: 9} is a plot of the maximum value 
of $K(\eta,s)$ along a line of constant $s$, for all $\eta>0$ . Because this maximum 
value is always negative, for all $s\in[0,1]$, the system is always dissipative, for all choices 
of $s$.

\begin{figure}
\centering
\includegraphics[width=0.5\textwidth]{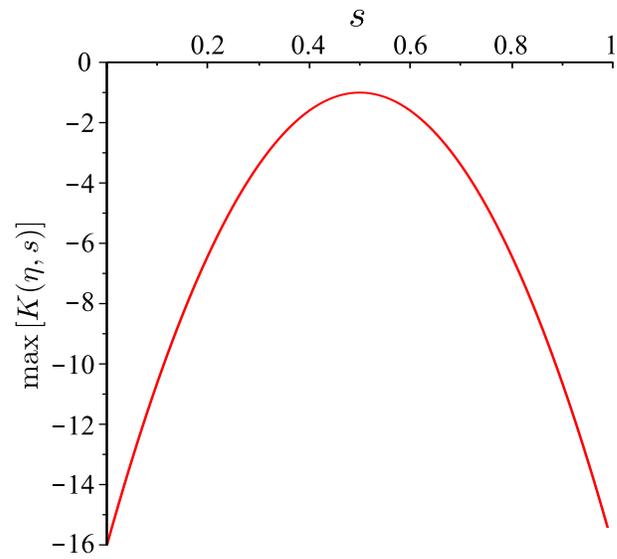}
\caption{
\label{fig: 9}
Plot of the maximum value of $K(\eta,s)$ (defined by (\ref{eq: B.3}) along a line
of constant $s$. The fact that this is always negative ensures that the centre of gravity 
is always sinking.}
\end{figure}


\begin{thebibliography}{99}

\expandafter\ifx\csname natexlab\endcsname\relax
\def\natexlab#1{#1}\fi
\expandafter\ifx\csname selectlanguage\endcsname\relax
\def\selectlanguage#1{\relax}\fi

\bibitem[Brenner (1961)]{Bre+61}
{\sc  Brenner, H. } 1961  
{The slow motion of a sphere through a viscous fluid towards a plane surface}. {\it Chem. Engng Sci.}  {\bf 16}, 242.

\bibitem[Cawthorn and Balmforth (2010)] {Caw+10}
{\sc  Cawthorn, C.J. and Balmforth, N. J. } 2010  {Contact in a viscous fluid. Part 1. A falling wedge}.
 {\it J. Fluid Mech.}  {\bf 646}, 327-38.


\bibitem[Happel and Brenner (1983)]{Hap+83}
{\sc  Happel, J. and Brenner, H. } 1983  
{Low Reynolds Number Hydrodynamics}. Martinus Nijhof, Hague, ISBN-13: 978-90-247-2877-0.

\bibitem[Michell (1950)]{Mic50}
{\sc Michell, A.G.M.} 1950  Lubrication, its Principles and Practice. Blackie.

\bibitem[Reynolds (1886)]{Rey86}
{\sc Reynolds, O.} 1886 {On the Theory of Lubrication and Its Application to Mr. 
Beauchamp Tower's Experiments, Including an Experimental Determination 
of the Viscosity of Olive Oil}. {\it Phil. Trans. Roy. Soc. Lond.} {\bf 11}, pp. 157-234.


\bibitem[Szeri (1998)]{Sze98}
{\sc  Szeri, A.S.}  1998 Fluid Film Lubrication. Cambridge University Press,  ISBN 978-0-521-89823-2.


\end{thebibliography}
\end{document}